\documentclass[screen,nonacm,10pt]{acmart}

\usepackage{ifthen}
\usepackage{xspace}
\usepackage{etoolbox}
\usepackage{cleveref}
\usepackage{paralist}
\usepackage{enumitem}
\usepackage{xcolor}
\usepackage{soul}

\newtoggle{comments}
\togglefalse{comments}

\newcommand{\participantquote}[2]{%
  \begin{quote}
    \emph{"#1"} ---#2
  \end{quote}%
}

\AtBeginDocument{%
  \providecommand\BibTeX{{%
    \normalfont B\kern-0.5em{\scshape i\kern-0.25em b}\kern-0.8em\TeX}}}

\graphicspath{{./images/}} 

\usepackage{tabularx}
\usepackage{array}

\begin{document}

\title{Two's a Crowd: Human and AI-Based Copresence for Developers with ADHD}

% Unblinded Author Details
\author{Veronica Pimenova}
\affiliation{%
  \institution{University of Michigan}
  \city{Ann Arbor}
  \state{Michigan}
  \country{USA}
}
\email{pimenova@umich.edu}

\author{Seth Bernstein}
\affiliation{%
  \institution{University of Michigan}
  \city{Ann Arbor}
  \state{Michigan}
  \country{USA}
}
\email{sethbern@umich.edu}

\author{Shalini Madan}
\affiliation{%
  \institution{University of Michigan}
  \city{Ann Arbor}
  \state{Michigan}
  \country{USA}
}
\email{shalinii@umich.edu}

\author{Dhruv Jain}
\affiliation{%
  \institution{University of Michigan}
  \city{Ann Arbor}
  \state{Michigan}
  \country{USA}
}
\email{profdj@umich.edu}

\author{Venkatesh Potluri}
\affiliation{%
  \institution{University of Michigan}
  \city{Ann Arbor}
  \state{Michigan}
  \country{USA}
}
\email{potluriv@umich.edu}

\renewcommand{\shortauthors}{Pimenova et al.}

\begin{abstract}
    Effective collaboration and communication are vital to developer productivity and well-being, yet remain constrained by human factors such as attention, intrinsic motivation, and interpersonal accountability. These constraints are particularly vital for developers identifying with Attention Deficit Hyperactivity Disorder (ADHD), who navigate persistent environmental barriers in modern hybrid workplace settings. While developers with ADHD frequently rely on collaborative copresence practices (such as body doubling or pair programming) to support executive function, the recent emergence of agentic AI coding assistants has begun reshaping these collaborative dynamics. To investigate how developers with ADHD engage in human and AI-based copresence practices, we conducted semi-structured interviews with 14 software engineers with ADHD. Our findings reveal that while traditional human-human copresence provides critical social support and onboarding structure, it forces developers to constantly manage professional reputation and sacrifice personal privacy. Conversely, developers leverage emerging human-AI copresence to maintain accountability and cognitive flow without the social anxiety, performance judgment, or surveillance associated with human observation. Based on these empirical insights, we map developer copresence practices onto core dimensions of Goffman's copresence theory and Forsgren et al.'s SPACE framework of developer productivity, and provide design recommendations for AI-based tools that promote inclusive collaboration for developers with ADHD.
\end{abstract}




\maketitle

\begin{figure}[htbp] 
  \centering
  \includegraphics[width=1\linewidth]{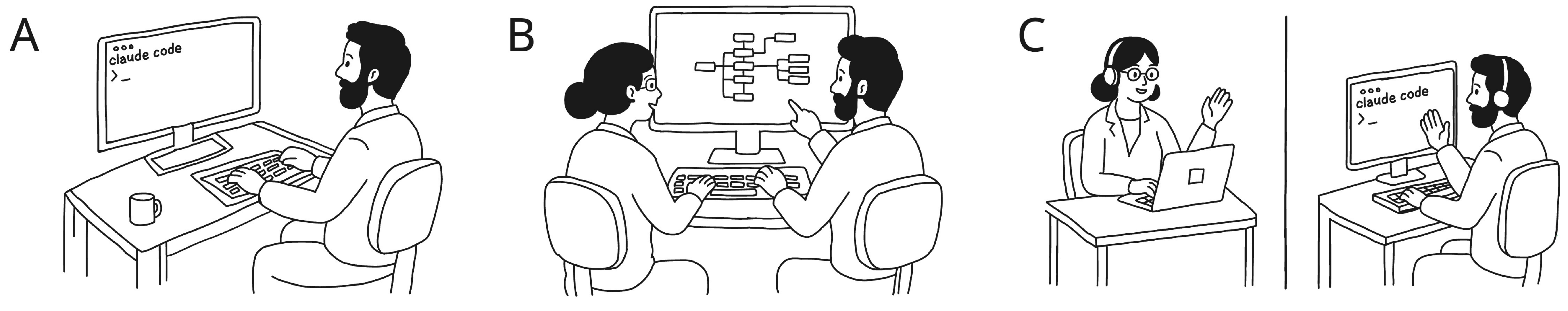} 

  \Description{Teaser image with three panels (drawing in black pen on a white background): A) is a developer sitting on a chair with a monitor that says "claude code", implying that the developer is working with an AI tool on their own with the AI acting as a copresence partner thus depicting human-AI copresence, B) is two developers sitting at a desk together in-person working at the same physical monitor setup and the monitor does not have AI on it, implying human-human pair programming or human-human copresence, and C) shows two developers working remotely at seperate desks in different locations, where one developer's screen is facing away from the viewer and the other developer's screen says "claude code" in text, implying that the second developer is using AI to write code and thus this image depicts human-AI-human copresence, where two humans are working together with AI assistance. This image shows A), B) and C) as three distinct copresence modalities with varying levels of AI use.
  }

  \captionsetup{justification=centering}
  \caption{Modalities of copresence practices for developers with ADHD: (A) human-AI copresence; (B) human-human copresence; (C) human-AI-human copresence.}
  \label{fig:intro-chart}
\end{figure}

\section{Introduction}
\label{sec:01-introduction}

Effective collaboration and communication among developers are vital to the productivity and success of software development teams~\cite{space_dev_productivity, evtikhiev2025what, bird_dev_productivity}. Traditional collaboration and communication patterns have changed rapidly with the rise of hybrid workplace practices and the widespread adoption of agentic, artificial intelligence (AI) software development tools~\cite{liu2025happening, ford_remote_swe}. In this evolving workplace, human factors such as sustained attention, intrinsic motivation, and interpersonal accountability have become critical determinants of overall developer productivity and well-being ~\cite{storey2026technical, treude2025generative}, as measured by Forsgren et al.'s SPACE framework for developer productivity (a widely-used metric in software engineering communities) ~\cite{space_dev_productivity}. 

Human factors are especially prevalent for the approximately 10.57\% of developers globally identifying with Attention Deficit Hyperactivity Disorder (ADHD)~\cite{stackoverflow2022}. While professionals with ADHD possess distinct cognitive strengths, including high levels of creativity and the ability to hyperfocus when working independently~\cite{boot2020creativity, White2011Creative, Stolte2022Characterizing}, these traits are frequently disrupted by workplace-induced environmental barriers. Unstructured workplace cues and fragmented, asynchronous communication forces developers with ADHD into continuous, exhausting self-regulation ~\cite{newman2025groove, das}. This ongoing environmental mismatch between cognitive needs and workplace demands contributes to lower job satisfaction, higher burnout, and increased rates of involuntary job termination ~\cite{Kessler2005ADHDWorkplace, Murphy1996ADHDAdults} 

To navigate these environmental barriers, professionals with ADHD frequently turn to \textbf{copresence}-based collaborative strategies such as body doubling in general daily contexts and pair programming within code development contexts~\cite{varnold, williams_pair}. Copresence involves leveraging the physical or virtual presence of another individual to enhance task initiation, self-regulation, and accountability during focus periods~\cite{CamposCastillo2013, Goffman1963}. In software engineering, pair programming similarly bridges executive function gaps by providing real-time cognitive support during complex tasks such as debugging or architecture planning~\cite{pair_expertise, newman2025groove}. As developers increasingly adopt agentic coding assistants as synthetic partners in AI-based co-creation~\cite{sarkar2025vibe, vibe_code_qual}, copresence is expanding beyond human-human dynamics to include human-AI collaboration. However, while traditional human-human copresence provides critical social-emotional and accountability support, it remains unclear how these dynamics translate when the copresence partner is an AI agent or how AI tools enhance human-human collaboration.

To address this gap, our study investigates the following research questions:

\begin{itemize}
\setlength{\itemsep}{1mm} 
    \item \textbf{RQ1}: How do software engineers with ADHD currently engage in copresence practices across human-human and human-AI collaboration settings?
    \item \textbf{RQ2}: What challenges do developers with ADHD experience during copresence practices, and what strategies do they employ to maintain focus and flow across human-human and human-AI copresence practices?
    
    \item \textbf{RQ3}: How can AI-based tools be designed to enhance copresence practices--both as partners in human-AI collaboration and as facilitators of human-human collaboration?
    
\end{itemize}

Through semi-structured interviews with 14 software engineers with ADHD, we uncover how developers navigate mutual copresence practices across physical, remote, and AI-mediated settings. We reveal that while in-person copresence practices such as body doubling provide social accountability and task initiation, they often introduce anxiety related to job performance. Conversely, while agentic AI tools offer judgment-free technical pairing and real-time feedback, uncoordinated AI usage increases cognitive load and disrupts shared mental models. Finally, we highlight how developers establish boundaries where AI agents succeed as ambient focus tools and where human expertise remains.

In summary, we contribute:

\begin{itemize}
\setlength{\itemsep}{1mm} 
    \item \textbf{Empirical findings} from semi-structured interviews with 14 developers with ADHD detailing how developers navigate executive dysfunction, task initiation barriers, and social anxiety during human-human and human-AI copresence sessions.
    
    \item \textbf{Theoretical mapping} of collaborative copresence practices, challenges and AI workflows onto copresence theory, structured around dimensions of Forsgren et al.'s SPACE framework. 

    \item \textbf{Design implications} for AI-based copresence tools that balance productivity with emotional and cognitive well-being for developers with ADHD.
\end{itemize}

\section{Background \& Related Work}
\label{sec:02-background}
We contextualize our work within four areas of prior literature: (1) ADHD in workplace settings, (2) the theoretical foundations of copresence and body doubling, (3) collaborative practices in software engineering, and (4) prior work on developers with ADHD.

\subsection{ADHD in Workplace Settings}
ADHD is a cognitive, neurodevelopmental condition in which individuals experience high levels of creativity, divergent thinking, and hyperfocus (a state of deep, immersive engagement with complex tasks)~\cite{Sedgwick2019, boot2020creativity, hupfeld2019living, Oroian2024Hyperfocus}. Beyond individual cognition, the differences between individuals with ADHD and neuro-typical individuals influence how professionals with ADHD navigate social and collaborative workplace settings~\cite{Barkleyetal2008, leveling_playing_field}. Workplace functioning for adults with ADHD is often shaped less by ability alone and more by organizational structure, where organizational demands (e.g. strict deadlines, multitasking, and low external support) can exacerbate executive function challenges, leading to disproportionate effects~\cite{nadeau2005career}. 

Adults with ADHD experience higher rates of unemployment and significant barriers in occupational functioning in traditional workplace settings, including increased job burnout characterized by higher emotional, cognitive, and physical exhaustion~\cite{kessler2006prevalence, turjeman2024executive}. These environmental constraints prevent professionals with ADHD from leveraging their strengths (such as creativity or hyperfocus~\cite{Sedgwick2019, boot2020creativity, hupfeld2019living}) in workplace settings. Following the \textit{Social Model of Disability} ~\cite{Goodley2012Disability, shakespeare2006social}, we frame these barriers not as individual deficits, but as the result of social and physical barriers created within traditional workplace environments 
~\cite{Barnes2019, baillargeon2025social}. 

A lack of structured workplace cues often forces professionals with ADHD to perform invisible access labor: the additional, unacknowledged mental effort required to adapt neurotypical workplace systems to meet their own cognitive needs~\cite{das, Wang2022Invisible, Branham2015Invisible}. Das et al. explored how in remote work environments, the "invisible" nature of ADHD can lead to a lack of structured workplace support, forcing professionals to perform \textit{invisible access labor} to maintain a balance between meeting performance expectations and managing their own mental energy~\cite{das}. Ezeamii et al. similarly find that PhD students with ADHD engage in invisible labor to navigate rigid academic structures by developing personalized workflows, relying on informal support networks, and avoiding stigmatized formal accommodations ~\cite{ezeamii2025navigating}. The invisible labor required by ADHD professionals in workplace settings creates significant friction in balancing productivity and well-being, where Pimenova et al. explored how professionals with ADHD experience strenuous invisible access labor and create their own "hacks" to assistive technology just to access everyday workplace communication systems~\cite{adhd_chronic_illness}. Marathe and Piper expand the concept of invisible access labor into the \textit{accessibility paradox}: while companies actively seek to hire and retain disabled workers, reliance on product-driven goals and normative productivity metrics leads to the de-prioritization of internal tool accessibility, forcing disabled employees to perform invisible labor ~\cite{marathe2025accessibility}.

Furthermore, the lack of physical copresence in digital spaces such as Zoom can increase invisible access labor, leading to isolation or workplace friction~\cite{zoom_auto, liu2025happening}. Duckert and Bj{\o}rn show that the spatial instability inherent in hybrid work introduces "location multiplicity"—an unpredictability in physical office attendance that causes workers to default to digital tools, turning physical office spaces into a lost space and compounding the difficulty of maintaining shared awareness~\cite{duckert2025location}. Conversely, Meyer and Fritz show that cultivating shared schedule awareness and unified presence displays in hybrid teams can mediate intrusive interruptions, allowing knowledge workers to better protect deep focus time while maintaining effective teamwork~\cite{meyer2025better}. These mismatches are especially prevalent in software engineering settings, where approximately 10.57\% of software engineers identify as having ADHD~\cite{stackoverflow2022, Verma2026}. Many developers with ADHD find that a lack of structured, physical workplace cues creates a reliance on internal self-regulation that is difficult to sustain~\cite{Fuermaier2021, newman2025groove, newman_disclosure}. This highlights a critical gap in the exploration of how collaborative practices in workplace settings can be changed to reduce \textit{invisible access labor} and promote well-being for developers with ADHD.

\subsection{Copresence and body doubling}
Previous literature in the field of social and behavioral psychology has defined the action of an individual "being together" with another individual as the term \textit{copresence} ~\cite{Goffman1963, Zhao2003}, which creates a sense of a shared environment for two individuals. Campos-Castillo and Hitlin breaks copresence into three primary components: mutual attention, mutual emotion, and mutual behavior~\cite{CamposCastillo2013}. Mutual attention refers to two individuals who are "reciprocally focused on one another", mutual emotion refers to the sharing of the other person's emotion through conscious awareness, and mutual behavior refers to a combined behavior pattern in which one person mimics another's motor activity~\cite{CamposCastillo2013}. This sense of "mutual togetherness" can be extended beyond physical presence to virtual environments~\cite{Schroeder2006, Survey_copresence, Bachmann2021, Subramaniam2013, tenenberg2016awareness}. This can happen in digital environments through virtual avatars, online Zoom calls, or asynchronous settings such as chat rooms ~\cite{Schroeder2002, kang_social_copresence_avatar, lee_ambient_copresence, goodwin_being_there, teruyama_spatial_copresence}. 

A specific manifestation of copresence which has gained significant traction within the ADHD community is \textit{body doubling}, where an individual performs a task in the presence of another person (the ``body doubling partner'')~\cite{ADDA2025, Born2024Effects}. This practice increases motivation for task completion through a shared sense of "working alongside" another~\cite{eagle_investigation, ADDA2025}. Eagle et al. explored how body doubling practices provide a non-judgmental accountability structure for individuals with broader neurodivergence, including ADHD~\cite{eagle_investigation, eagle_body_doubling_og}. Arnold et al. demonstrated the effectiveness of body doubling for students in post-secondary education as a tool to manage academic load~\cite{varnold}. In software engineering contexts, where tasks such as debugging involve extended periods of cognitive demand, these mechanisms may be particularly valuable~\cite{space_dev_productivity}. However, while high-intensity collaborative practices such as pair programming are well-documented~\cite{begel_pair}, less is known about how "low-intensity" forms of copresence, such as body doubling or virtual coworking, are leveraged by professional developers with ADHD to navigate their daily tasks and workflow. We contribute empirical insights into how developers with ADHD leverage low-intensity copresence modalities (body doubling and pair programming) to manage executive function demands and support daily workflows.

\subsection{Collaboration in software engineering settings}

Human factors of software engineering (e.g. communication, social well-being) have increasingly become more recognized as vital to developer productivity over technical factors (such as code output or technical performance) alone~\cite{space_dev_productivity}. We follow Forsgren et al.'s \textit{SPACE} framework of developer productivity, which includes \textbf{S}atisfaction and well-being, \textbf{P}erformance, \textbf{A}ctivity, \textbf{C}ommunication and collaboration, and \textbf{E}fficiency and flow ~\cite{space_dev_productivity}. The SPACE framework is a multidimensional model for measuring developer productivity across five key dimensions, synthesizing previous literature across fields, including areas of human-computer interaction, software engineering, and organizational psychology~\cite{space_dev_productivity}. Central to this framework is the concept of "flow", a state of uninterrupted, deep focus that is essential for complex programming tasks~\cite{space_dev_productivity}. 

Further, individual developer productivity and well-being is embedded within team-level collaboration and the social environment in which code is produced~\cite{evtikhiev2025what, jackson2022collaboration}. Devathasan et al.
describe that while diversity in software engineering teams enhances creativity 
and performance, teams must first have the ability to overcome hardship and empathy~\cite{Devathasan2025Empathy}. Evtikhiev et al. further emphasize that software development collaboration is often shaped by breakdowns in communication, coordination, and knowledge sharing that can accumulate over time and negatively impact team effectiveness ~\cite{evtikhiev2025what}. Further, Miller et al. found that remote work disrupted communication ease and reduced opportunities for informal social interaction ~\cite{miller2021weekend}.

One of the most established collaborative practices in software engineering is pair programming, where two developers work synchronously on a single task~\cite{williams_pair, begel_pair, pair_expertise}. Begel et al. described that benefits of pair programming include fewer bugs, higher code understanding, and higher code quality~\cite{begel_pair}. Recently collaboration in software engineering has begun to expand beyond human partners to include AI-based tools that directly participate in the coding process, with platforms like GitHub Copilot or ChatGPT reaching over 15 million developers and 700 million weekly active users respectively ~\cite{deming2025how}. With this recent increased usage of AI in programming and software engineering contexts (e.g. "vibe coding")~\cite{sarkar2025vibe, vibe_code_qual}, there is an opportunity for developer support by pair programming with AI. Unlike human partners, AI agents scale effortlessly, are always available, and provide a non-judgmental environment for experimentation. This absence of social pressure is further reflected in how human-AI interaction dynamics evolve over time, where Lazebnik et al. found that social politeness norms erode significantly faster in human-AI interactions compared to human-human collaboration~\cite{lazebnik2025mind}. This human-AI interaction creates a new paradigm of ``AI-pair programming'' where the AI can provide a form of digital copresence~\cite{ma2023ai, zhou2025exploring, lyu2025will, spinellis2024pair}. 

Recent CSCW and HCI literature highlights how human-AI co-creation shifts 
collaboration patterns, where managing agency and control mechanisms is becoming a more important measure for how effectively users co-create alongside AI systems~\cite{zhang_human_AI_in_cscw, holter2024deconstructing}. Effective human-AI collaboration relies heavily on delegation behavior, where Spitzer et al. demonstrated that providing clear contextual information about both human and AI capabilities significantly improves team performance and optimizes task delegation~\cite{spitzer2025human}. While prior work has extensively studied the technical accuracy of AI-generated code, there is a gap in understanding the social and psychological impact of AI as a copresence partner, especially for those who find the social demands of human-to-human pair programming overwhelming (such as developers with ADHD).

\subsection{Prior work on developers with ADHD}
Recent empirical research has begun to explore the experiences of software engineers with ADHD. Newman et al. explored disclosure in workplace settings through a social media analysis and a large-scale survey of software engineers with ADHD and neurodivergence more broadly,  discovering both positive (workplace and social support) and negative (workplace stigma, discrimination, extra labor) outcomes from disclosure ~\cite{newman_disclosure}. Newman et al. also conducted a further social media analysis and survey specific to professional programmers with ADHD and found that body doubling was used to aid time management, consistent performance, and task initiation and completion specifically ~\cite{newman2025groove}.

Liebel et al. interviewed software engineers with ADHD and found challenges with task deadlines, organization and planning, and over-promising ~\cite{liebel_swe_adhd}. This affected developers physical and mental health, as well as task completion and workplace productivity~\cite{liebel_swe_adhd}. Strengths of developers with ADHD included puzzle solving, creativity and divergent thinking, and the ability to "think ahead", which increased workplace reward and task focus~\cite{liebel_swe_adhd}. Gama et al. also explored neurodivergent software engineers more broadly, interviewing four developers with ADHD on the emotional and social factors of being neurodivergent and resulting workplace adaptations and accommodations~\cite{gama_agile_teams}.

Despite the increasing body of work identifying challenges and strengths of software engineers with ADHD, research into collaborative interventions specifically tailored for developers with ADHD remains limited. While Newman et al.~\cite{newman2025groove} suggests that copresence practices, such as body doubling or pair programming, could theoretically scaffold the cognitive needs of this population, there is a lack of empirical data on how these practices are utilized in modern AI-based real-world software engineering settings. While we understand that human factors are vital to productivity~\cite{space_dev_productivity}, we do not yet know how developers with ADHD experience "low-intensity" copresence or how they might leverage AI agents to fulfill the role of a body doubling partner. Our study addresses this gap by exploring the intersection of varying ADHD cognitive styles, copresence theory, and modern AI-mediated workflows to promote software engineering workplace environments that are inclusive and enable productivity while ensuring developer wellbeing.

\section{Study Procedure}
\label{sec:03-methods}
We describe our study procedure including participant recruitment, interview procedure, participant demographics, and data analysis.

\subsection{Participant recruitment}
To gain deeper insights into the experiences of developers with ADHD related to copresence practices, we conducted 14 semi-structured interviews. To be eligible for our study, participants had to be professional software developers working in industry, have experience working on a team in an industry setting, self-identify with or be formally diagnosed with ADHD, and located in the United States. Participants were recruited through a variety of social media posts on platforms such as LinkedIn and X (Twitter). Participants completed a pre-interview recruitment survey, where they described their programming experience and current role. Additionally,  participants were required to indicate that they either have a formal ADHD diagnosis or self-identify with ADHD and score 14 or higher on the World Health Organization Adult ADHD Self-Report Scale (ASRS) for DSM-5~\cite{Born2024Effects}. We administered this scale as a screening instrument because fewer than 20\% of adults with ADHD are accurately diagnosed and treated ~\cite{RivasVazquez2023Adult}, due to the barriers to diagnosis in adult populations~\cite{Sgro2025Barriers}. The scale is able to screen participants, allowing us to provide a consistent inclusion criterion for participants without a formal diagnosis and correctly identifies 91.4\% of adults with ADHD and 96\% without ~\cite{kessler2006prevalence}.

Our final sample included individuals with 1 -- 25 years of industry experience in software development with in-person, hybrid, and remote workplace modalities, and with 11 participants identifying as male and 3 participants identifying as female (see  Table~\ref{tab:demographics} for the full participant demographic). Interviewees were compensated with a \$50 USD gift card and data collection was concluded when our sample provided sufficient depth and diversity of perspectives to comprehensively address our research questions.

\subsection{Procedure}
We conducted a semi-structured interview in three parts, structured around our guiding research questions. Part one asked questions about how developers currently engage in copresence practices, specifically focusing on body doubling and pair programming. In part two, participants shared insights into their experiences with these practices, including the specific benefits and challenges they encounter in industry workflows. Part three elicited responses about how an AI tool could act as a copresence partner or assist in the facilitation of copresence sessions.\footnote{We have attempted to make the study fully replicable. A replication package containing all necessary materials, including interview protocol and final, full codebook are available in supplementary materials of this submission. We will publicly share these files in a Zenodo link upon acceptance.} We performed three pilot interviews to finalize our interview protocol script. Interviews were conducted via Zoom, lasted approximately 60 minutes, and were audio-video recorded using Zoom's recording software. Following each session, the lead researcher created reflective memos, and the transcripts generated by Zoom's automatic captioning service were manually reviewed and corrected by the researcher.

\begin{table}[ht]
\centering
\caption{Participant demographic data of software engineers with ADHD ($N=14$)\label{tab:demographics}}
\small 
\begin{tabular}{lllllll}
\toprule
\textbf{ID} & \textbf{Years in Industry} & \textbf{Workplace Modality} & \textbf{Gender} & \textbf{Age} & \textbf{ADHD Diagnosis} & \textbf{ASRS Score}\\
\midrule
P1  & 1 year   & Hybrid  & Male         & 24  & Self-Diagnosis & 15\\
P2  & 5 years  & Remote & Male        & 27 & Self-Diagnosis & 19\\
P3  & 2 years  & Remote & Male        & 22 & Formal Diagnosis & 18\\
P4  & 2 years  & Hybrid & Male        & 23 & Self-Diagnosis & 16 \\
P5  & 2 years  & In-person & Male     & 28 & Formal Diagnosis & 15\\
P6  & 3 years  & Hybrid & Female      & 24 & Formal Diagnosis & 18\\
P7  & 1 year   & Hybrid & Male         & 23 & Self-Diagnosis & 23\\
P8  & 6 years  & Hybrid & Male        & 27 & Formal Diagnosis & 15\\
P9  & 2 years  & Remote & Male        & 27 & Formal Diagnosis & 17\\
P10 & 2 years  & In-person & Male     & 21 & Self-Diagnosis & 15\\
P11 & 4 years  & Remote & Male         & 24 & Self-Diagnosis & 16\\
P12 & 25 years  & Remote & Male         & 51 & Formal Diagnosis &  17\\
P13 & 1 year  & Hybrid & Female         & 22 & Formal Diagnosis &  19\\
P14 & 7 years   & Hybrid & Female   & 31 & Formal Diagnosis &   17\\
\bottomrule
\end{tabular}
\end{table}

\subsection{Data Analysis}
\begin{figure}[h] 
  \centering
  \includegraphics[width=1\linewidth]{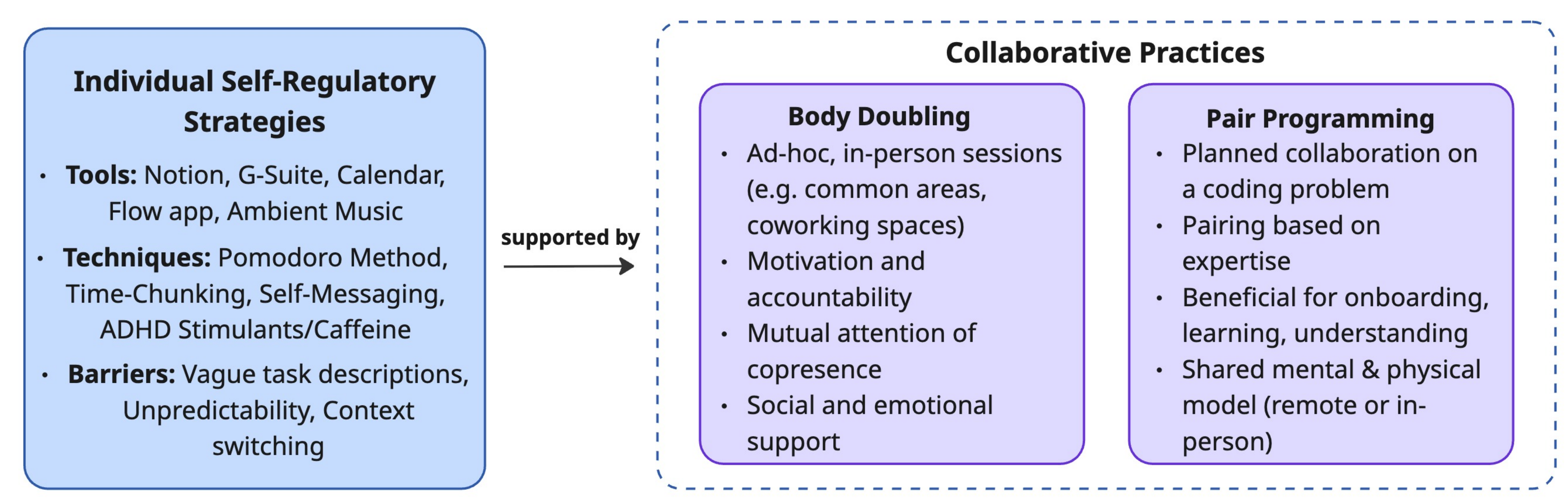} 

  \Description{A flowchart titled "Individual and Collaborative Workflows for Software Engineers with ADHD." On the left, a box labeled "Individual Self-Regulatory Strategies" lists tools (Notion, G-Suite, Calendar, Flow app, and Ambient music), techniques (Pomodoro method, time-chunking, self-messaging, ADHD stimulants/caffeine), and barriers (vague task descriptions, unpredictability, context switching). An arrow labeled "supported by" points to a larger dashed box on the right titled "Collaborative Practices." This box contains two sub-sections: "Body Doubling," which emphasizes ad-hoc, in-person sessions (e.g. common areas or coworking spaces) for motivation and accountability, mutual attention of copresence theory, and social-emotional support; and "Pair Programming," which describes planned collaboration on a coding problem, junior \& senior developer pairs, beneficial for onboarding, learning, understanding, and shared mental \& physical model (remote or in-person).
  }

  \captionsetup{justification=centering}
  \caption{Map of individual ADHD-related self regulatory strategies supported by collaborative practices}
  \label{fig:findings-chart}
\end{figure}

To analyze our qualitative data, we followed Deterding and Waters’ flexible approach to thematic analysis~\cite{Deterding2021Flexible}, which provides an iterative, top-down approach for semi-structured interview data. This is similar to the qualitative analysis process in related software engineering 
work ~\cite{newman2025groove, newman_disclosure}, where this process balances pre-existing research goals with inductive coding and allowed us to begin with broad, conceptual indexing mapped directly to our research questions (e.g., barriers to current copresence practices, aspects that could be improved with AI), while remaining open to nuances within developer workflows. Our analysis progressed through three stages: (1) Open coding: The 14 anonymized interview transcripts were randomly divided among three members of the research team (the primary author and two collaborators) after the first author created a set of initial codes based on existing research questions. Team members independently reviewed their assigned transcripts during an initial open-coding phase to create original, descriptive codes capturing participant behaviors and responses; (2) Visual mapping and sub-coding: To synthesize these initial codes, the team engaged in continuous analytic memoing and met twice over a four-week period to collaboratively group codes. We used Miro display boards to visually organize the boundaries of our sub-codes (as illustrated in Figure~\ref{fig:findings-chart}); (3) Final code refinement: The axial coding phase allowed us to condense our original codes into 23 distinct codes, organized into three key themes with 7--8 sub-codes per theme, ensuring sufficient empirical detail to capture nuances of participant workflows. Following Padiyath and Nelson-Fromm ~\cite{aadarsh}, we did not calculate Inter-Rater Reliability (IRR), as IRR does not align with the reflexive, interpretivist approach to thematic analysis used, which prioritizes organic, collaborative depth over enforced consensus.

\section{Findings}
\label{sec:04-results}

\subsection{Modality of current copresence practices}
\label{subsec:modality_copresence}

We describe the modality of current copresence practices employed by developers with ADHD through distinct structural modalities (\S\ref{subsubsec:distinct_structural_modalities}), motivating factors contributing to developer participation in such practices (\S\ref{subsubsec:motivational_mechanisms}), and the interpersonal partner dynamics that influence collaboration (\S\ref{subsubsec:interpersonal_partner_dynamics}).

\subsubsection{Distinct structural modalities}
\label{subsubsec:distinct_structural_modalities}

Copresence practices are widely used in ADHD communities to support task initiation and completion~\cite{varnold, eagle_investigation, eagle_body_doubling_og}, and our participants (as developers with ADHD)\footnote{From this point onward, when referring to "developers" or "software engineers" we refer explicitly to developers or software engineers with ADHD. When referring to developers or software engineers without ADHD, we explicitly use the word "neurotypical".} frequently relied on copresence to provide the structural support that individual planning tools (e.g. Google Calendar, Notion, personal messaging reminders) lacked.

However, participants described company-imposed environmental barriers such as unpredictable task volumes, unclear expectations, and requirements of continuous availability frequently disrupted personal planning and heightened performance anxiety. For example, P1 described feeling bound to their machine due to the immediate, unannounced assignment of tasks, creating an environment of constant pressure without adequate support systems for task initiation. Similarly, P12 highlighted the psychological weight of on-call obligations, explaining that continuous alert notifications often trigger avoidance behaviors such as snoozing notifications to manage the compounding stress of perpetual availability. P14 noted that corporate complexity further complicates personal estimation of task completion time, forcing them to manually pad schedules by several days to account for unexpected enterprise-based hurdles:

\participantquote{Planning is a process where I often undershoot things. I've gotten better at calculating times, but you never know what [software] you're going to be able to use or not use, or if you're gonna have to jump some hurdles just because you're in this gigantic enterprise environment. Usually I still need to add a day or two to whatever I have planned, because of unforeseen things. I've gotten better at it, but it's been challenging.}{P14}

To counter these barriers, developers turned to structured collaboration through copresence, often combining body doubling with Pomodoro-styled time cycles~\cite{cirillo2018pomodoro}.
These structured intervals provided external time management while reducing ADHD-specific cognitive fatigue and cognitive load ~\cite{yamamoto2022relationship}. Specifically, shared breaks allowed developers to mentally distance themselves from complex problems or cognitively overwhelming tasks. P14 described how their daily usage of body doubling with a family member through Pomodoro techniques allowed them to cognitively rest:

\participantquote{My sister and I use Pomodoro, so it's basically the 25 minutes and the 5-minute break. In the 5-minute break, it's usually catching up on "Oh, how was it?" It clears up the brain from what you're doing, so you're not still thinking about the problem that you're solving while you're on break.}{P14}

Participants also distinguished between physical and virtual closeness. While ad-hoc, in-person sessions were preferred for social-emotional support, participants favored remote, timed sessions when prioritizing task productivity and code quality. P5 used remote body doubling via synchronized Pomodoro schedules, where both partners wore headphones during quiet focus blocks and restricted conversation strictly to shared breaks.

While body doubling provides a baseline of accountability when partners perform different tasks, our participants utilized pair programming as a more intensive form of external executive function support for complex, cognitively demanding tasks. Unlike body doubling, where participants work on separate tasks, pair programming involves a high degree of mutual behavior where a partner’s actions directly influence the developer's next step. Participants (P1, P2, P3, P4, P5, P8, P9, P10, P11, P13, P14) emphasized that partners should explicitly align and make a concrete, shared plan together before typing any code, establishing clear guardrails for the session or sharing a single monitor in person to force mutual attention on a single line of code at a time. P3 executed pair programming sessions around structured plans: screen-sharing to code in small blocks, discussing each iteration, and updating documentation together to reinforce learning and maintain clear guardrails.

By splitting the system knowledge with a partner, participants are able to focus on specific code changes without losing track of the larger architecture in their codebase. To mitigate distractions that often occur when passively observing another individual's work, participants utilized a swapping strategy. By first investigating separate modules of a complex system independently, the developer pair could then act as guides for one another. P2 described this swapping technique as a way to maintain active engagement and manage the cognitive load of a new codebase: 

\participantquote{[We] spend some time separately learning the system, and then pair program a change together. He [my partner] knew half of the system, I knew the other half. I’d pair program and watch him do his changes, and then we can swap in and out.}{P2}

\subsubsection{Motivational mechanisms}
\label{subsubsec:motivational_mechanisms}

Participant engagement in copresence was primarily driven by collective momentum, externalized accountability, and real-time verbal validation.

Participants drew baseline motivation from the concept of mutual behavior, engaging in independent tasks alongside others in a shared environment. Consistent with previous literature on body double for individuals with ADHD~\cite{varnold, eagle_body_doubling_og, eagle_investigation}, simply observing peers or strangers working productively (e.g. in coffee shops or shared remote calls) transformed isolated, daunting tasks into a seemed shared effort. This is particularly relevant for remote workers, where P1 and P2 work primarily remotely and frequently body double with friends or strangers in coffee shops. P2 describes how simply knowing that others around them are also being productive is helpful for task initiation:

\participantquote{Working remote, sometimes I'm at my desk alone and it kind of sucks. So I would go to coffee shops or a couple other friends who work remote will get together. And just working in a spot where you know other people are working does help. I just have a feeling they're doing something too.}{P2}

This ambient presence created a sense of social pressure that lowered the threshold for task initiation and helped sustain focus without requiring direct interaction.

Beyond motivation, our participants also expressed that body doubling provided externalized \textbf{accountability}. Participants define the practice as two people trying to keep each other accountable while working on their own task, relying on the presence of a peer to maintain focus on complex tasks. During set focus time, participants wanted body doubling partners who would check-in on them or offer task support when needed. These focus sessions fulfill the mutual emotion component of copresence by providing social-emotional support through check-ins and structured breaks. P5 noted the balance between productivity and well-being gained from social interaction: 

\participantquote{I associate it [body doubling] with productivity and I don't associate it with anything negative. I think [it is] accountability, and then also during breaks, like, being able to talk to friends. I think it is a very nice break. So mainly those two aspects.}{P5}

Participants such as P8 also noted that while social interaction with their copresence partner provided social-emotional support, it should be time-bound through Pomodoro-style timers (as mentioned in \S\ref{subsubsec:distinct_structural_modalities}). Without strict timing, social interaction could create disruption and decrease productivity of a copresence session.

Finally, when participants shifted from passive body doubling to active collaboration (e.g. pair programming), the motivational mechanism evolved from accountability to cognitive validation. P2 noted that while they occasionally lose focus during passive observation, being an active participant in a discussion provided social accountability needed to stay engaged. By externalizing their thought process to a partner, participants move from a state of internal confusion to active problem-solving. P2 and P3 both described how pair programming also provides validation through real-time verbal feedback that reduces performance anxiety and catches obvious mistakes they may have missed on their own.

\subsubsection{Interpersonal partner dynamics}
\label{subsubsec:interpersonal_partner_dynamics}

The interpersonal success of a copresence session is heavily dependent on the underlying relationship dynamics. Participants emphasized that effective collaboration required psychological safety and a sense of mutual respect, which increased satisfaction and well-being during collaboration. Conversely, a lack of trust or safety heightened performance anxiety and reduced satisfaction. P8 describes comfort with engaging in focus sessions with a partner they trust:

\participantquote{Whether I can focus largely depends on the person I'm with. Co-workers that I like to be around, and that I trust to actually have good input keep me on task. People who I trust more and respect their abilities more, it's easier for us to collaborate.}{P8}

While some developers accepted strangers for passive body doubling, most preferred familiar partners who could balance quiet focus blocks with friendly conversation and check-ins during breaks for social-emotional support. P2 noted that an ideal partner should be sociable and non-threatening:

\participantquote{I guess someone having knowledge in the area is nice, if I do really hit a point I'm stuck, I can just quickly talk to them about it. It would be really weird if they're antagonistic to me. That just freaks me out.}{P2}

When selecting a pair programming partner, developers intentionally sought individuals with different or complementary skills. Partnering with a more experienced peer with complementary skills  allowed participants to overcome task-initiation hurdles and maintain a consistent mental model of complex codebases. For example, P3 compared their pair programming sessions to an "apprenticeship" that enabled them to learn complex systems by active participation rather than reading passive documentation. P9 described how complementary skills assisted with learning during pair programming:

\participantquote{Pair programming was helpful as a learning tool. It is the equivalent of having an LLM trying to fix your code with you instead of fixing it for you. So it was helpful as somebody who's new and trying to learn. My partner had the expertise to know where to put stuff, and I knew what needed to be put in there, so it felt complementary.}{P9}

\subsection{Challenges and Mitigation Strategies of Current Copresence Practices}
\label{subsec:challenges_strategies_copresence}

While copresence offers essential structural support for developers with ADHD, it also introduces performance anxiety (\S\ref{subsubsec:social_evaluative_anxiety}), disrupts flow state when executed poorly (\S\ref{subsubsec:flow_comprehension_tradeoffs}), and does not account for company-related privacy constraints (\S\ref{subsubsec:privacy_remote_tradeoffs}).

\subsubsection{Performance Anxiety}
\label{subsubsec:social_evaluative_anxiety}

Even before collaboration begins, executive dysfunction and emotional task avoidance raise barriers to task initiation, creating challenges around scheduling, planning, and interpersonal coordination required to initiate a copresence session. P6 described wanting to utilize body doubling, but routinely struggled to initiate sessions due to internalized anxiety--a sentiment echoed by P12 who valued body doubling but hesitated to reach out to their peers:

\participantquote{I like the other members of my team, but I don't want them to watch me work or bother them. I think it goes back to when I was young, and I had a feeling that I wasn't doing things the way I was supposed to do them and a feeling that other people are watching me, that I'm kind of doing things wrong in some way. It's quite deep.}{P12}

This fear of observation extends into live collaboration and is heightened by the need of participants to maintain a professional persons around colleagues. This dynamic created two challenges: it made initiating body doubling sessions intimidating, and once in a session, an internalized fear of asking clarifying questions hindered active problem-solving. P13 described how overcoming this anxiety to ask clarifying questions is vital for preventing systemic mistakes. Framing partner pairings  as a reciprocal, problem-focused partnership can help decrease performance anxiety.

\subsubsection{Disruption to Flow State}
\label{subsubsec:flow_comprehension_tradeoffs}

Beyond emotional overhead, real-time collaboration presents distinct cognitive challenges for our participants. An example is when partners are not synchronized in focus, which can create distractions. P6 described this as a lack of "harmonic co-working." For developers who are in a state of hyperfocus, live pair programming on smaller tasks can disrupt their internal map of the codebase, where P10 only wanted ambient music over human presence for deep focus:

\participantquote{I wouldn't use pair programming for any deep coding that I need to be in a flow state for. The only distraction I could have is minimal music.}{P10}

Unpredictable workplace interruptions also disrupt flow, where P5 and P1 noted avoiding copresence altogether during on-call periods due to the guilt of disturbing partners or losing momentum during sudden context shifts. P1 explained how unscheduled calls prevent them from participating in copresence practices alltogether:

\participantquote{It's very common to get calls out of the blue. You could be working on something for 3 hours, then all of a sudden, boom, get a call with a new thing, forget everything from before. I like to go and body double in public, but then I don't know when I'm gonna take a call. I don't want to distract others and I don't want to lose my own focus.}{P1}

\subsubsection{Privacy Constraints}
\label{subsubsec:privacy_remote_tradeoffs}

Finally, the transition to remote collaboration introduces structural trade-offs between security policies and executive function support. P6 and P9 emphasized that corporate NDAs and data privacy regulations severely restrict traditional screen-sharing practices. Our participants relied on screen sharing during body doubling and pair programming sessions for visual accountability, where they would risk sharing sensitive information to outsiders or feel a sense of heightened anxiety when screen sharing with coworkers. P6 described the need for a privacy-preserving alternative towards screen-sharing during copresence sessions:

\participantquote{Sometimes I am just browsing information and accidentally lose control [of my focus], but I don't want to share my full screen in front of everyone. I'm wondering if there is a design that could simulate in-person collaboration such as a neighbor who can see my screen, but cannot see what exactly I am doing. That way of accountability would be more helpful and private.}{P6}

\subsection{The Role of AI in Copresence}
\label{subsec:AI_in_copresence}

While participants were largely open to AI as a copresence partner, they expressed nuanced perspectives regarding how AI can enhance productivity, where it falls short, and how it compares to human interaction. We describe how AI supports flow state (\S\ref{subsubsec:ai_executing_copresence}), human boundaries in AI-mediated copresence (\S\ref{subsubsec:ai_task_boundaries_expertise}), and social-emotional support (\S\ref{subsubsec:humanity_tension}).

\subsubsection{AI Supports Flow State}
\label{subsubsec:ai_executing_copresence}

The usage of agentic coding tools (predominantly Claude Code and GitHub Copilot) was frequent and widespread, where all participants had experience with AI for programming and most participants (P1, P3, P5, P6, P7, P8, P9, P10, P12, P13) had positive perceptions. Out of participants who had negative perceptions (P2, P4, P11, P14), only one participant (P11) completely avoided AI use. Generally, participants with positive perception of AI characterized the AI tools as pair programming partners, who could create uninterrupted, long periods of flow state. P5 describes:

\participantquote{It's almost like a honeymoon and I actually cannot see any project that I would not want to use it for. Ironically, I probably have more positive feelings towards pair programming with an AI than with a human... because unlike a human, [the AI] doesn't interrupt my flow with unrelated small talk that distracts or hinders moving forward.}{P5}

Further, agentic AI workflows support multi-stream execution, allowing developers to manage parallel streams of problem-solving across interfaces without losing cognitive context. For instance, P10 described rapidly orchestrating work across terminals, notes, GitHub, and local code bases, leveraging concurrent AI agents to maintain momentum without getting overwhelmed in context switching:
\participantquote{I have two different terminal apps. I just opened Claude inside of the terminal, then I'll have, Obsidian. I'll have my codebase, in Finder next to it, and I'll have another tab open with GitHub and another one open with my notes. And then I'll jump back and forth and I might open up a new agent. A lot of orchestrating things because it's a lot faster and each stream of thought updates based on what you're reading from each tab. It's actually kind of goated.}{P10}

AI-based tools also offload additional workload by digesting large amounts of documentation and providing code validation in real time. Rather than relying on AI to write end-to-end solutions, participants often used AI to establish high-level skeletons before stepping in to review and refine the output. For instance, P7 used AI agents to summarize complex codebases and generate initial structural frameworks, allowing human effort only for auditing output. By continuously prompting developers for clarification and providing task summaries, the AI tools forced developers to externalize, articulate, and organize their thought processes.

\subsubsection{Human boundaries for AI copresence}
\label{subsubsec:ai_task_boundaries_expertise}

Despite the AI-mediated copresence providing support with entering flow state, our participants drew boundaries where human support is still needed. When code validation or domain expertise is required for complex or sensitive projects, developers experience heightened accountability-based anxiety. Due to AI tools bearing no systemic accountability, using an AI tool as a copresence partner (particularly for pair programming) shifts cognitive burden of auditing code and searching for errors onto the individual developer. To mitigate this stress, participants prioritized human copresence for trusted code verification. P5 noted:

\participantquote{I actually really like working with AI agents, but in terms of human pair programming, a really good use case would be where the other person has more expertise than I do with a particular thing or where I genuinely cannot tell if the code I've written is accomplishing the conceptual task. In those scenarios, it's very helpful to have someone [a human] there for verification.}{P5}

This boundary becomes especially pronounced in sensitive projects or high-stakes production environments, where participants described the process of manually auditing large volumed of AI code increases cognitive exhaustion and anxiety over hidden bugs. P11 emphasized that using AI for pair programming without human verification created individual cognitive burden:

\participantquote{Infrastructure and production stuff never gets generated for me. You can't afford a failure in these mission-critical systems. It kind of turns into spaghetti after a couple weeks. Everyone's like, "Oh, but I can do this with Claude in two hours." I'm like, yes, but can you maintain it? Can you scale it? Then if it doesn't work it's all on me.}{P11}

Similarly, P10 explained that without high-level architectural oversight, rapid AI code generation risks optimizing execution speed at the expense of strategic direction:

\participantquote{I use the analogy where you're in super-fast car... This is AI agents, you have no bird's eye view to tell you if you're going the right way. A human would ask you, 'Why are you going this way?'}{P10}

Further, when human-human copresence sessions include both partners using individual AI tools, rapid and unannounced AI usage by one partner can hinder the pair's shared mental model. In these hybrid sessions, participants expressed cognitive overload when a partner offloaded work to AI agents in real time without transparent communication. This challenge highlights how unaligned AI usage disrupts mutual attention--when one developer independently prompts an AI without keeping their partner informed. P14 detailed this breakdown:

\participantquote{It ended up being pair-programming without us actually wanting it to be. Mostly it was [my partner] telling Claude a bunch of stuff, and Claude doing a bunch of stuff, and me saying, "What is happening?" It was stressing, because I was trying to keep up with what the AI is doing and also you have the other person that is quickly changing screens.}{P14}

\subsubsection{Social-emotional support}
\label{subsubsec:humanity_tension}

A central trade-off in AI-mediated copresence sessions involves balancing social accountability with authentic emotional support. 
Our participants noted that when AI acts as a copresence partner, performance anxiety related to human observation is lowered. However, participants described how AI tools lack social support and emotional connection, meaning AI-based copresence lacks a sense of mutual emotion.\footnote{Mutual emotion is a core component of copresence theory~\cite{CamposCastillo2013}.} P2 explained how simulated social interactions with AI do not provide social-emotional support, as AI agents lack human emotion:

\participantquote{I don't feel like an AI gets bored. I can relate to someone around me who's just like, you know what, I need a break for a minute. Let's talk or something. If an AI did it, it's clearly doing it just to placate me.}{P2}

In some cases, the absence of human emotional traits can increase a developer's productivity. An AI-based copresence partner offers a predictable and controllable environment, allowing developers to enter flow state without social anxiety. P10 explains how social and emotional support are less important for success in their role, where they primarily aim to increase their everyday productivity:

\participantquote{I think the social or emotional support are less relevant. I think people or friends should follow that purpose. I think [for AI] productivity is better. If you have a set of tasks that are automatable, just ask an AI to write a script for you. I don't want to befriend it.}{P10}

Additionally, AI-based copresence partners can assist with structuring routine breaks that alleviate cognitive load. Beyond executing technical tasks, agents can simulate light, human-like check-ins which provide visual and temporal transitions during focus periods. P6 explains how an AI tool could mimic human interaction through check-ins:

\participantquote{If [An AI] periodically checks in whether I need to take a break and grab a coffee sounds more human. Sometimes when [a human colleague] checks in on me, that 5-minute talk is really helpful for stress relief.}{P6}

While our participants generally recognized that current AI agents couldn't fulfill the social-emotional support gained from human interaction, some participants either didn't want human interaction during focus periods, or have suggestions for how AI tools could seem more human. This shows potential for the development of AI tools that could act as copresence partners, which we discuss in \S\ref{subsec:impact_assistive_tech}.
\section{Discussion}
\label{sec:05-discussion}

\begin{figure}[h] 
  \centering
  \includegraphics[width=1\linewidth]{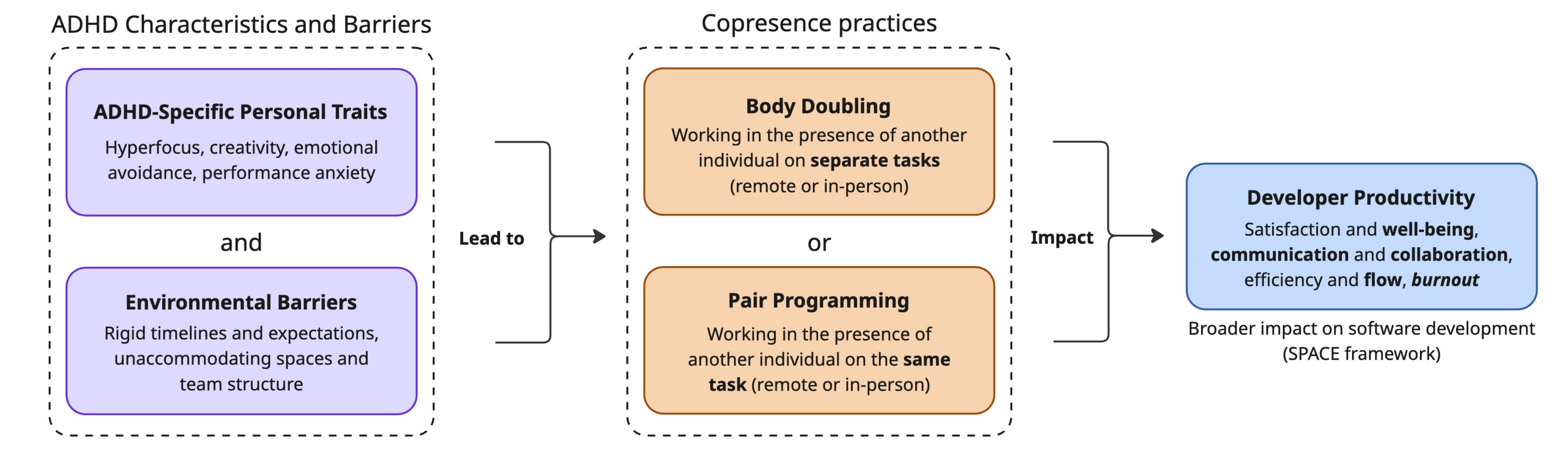} 

  \Description{Flow chart, from left to right starts with a dotted box titled "ADHD Characteristics and Barriers" which has two boxes inside of it, on top is "ADHD-specific personal traits" including the text "hyperfocus, creativity, emotional avoidance, performance anxiety". Then there is the word "and" which is followed by a box below titled "Environmental Barriers" and includes "rigid timelines and expectations, unaccommodating spaces and team structure". Then there is the text "lead to" and an arrow pointing to another two sets of boxes titled "Copresence practices" with the first, top box titled "body doubling" includes "working in the presence of another individual on seperate tasks (remote or in person)", then there is the word "or" which is followed by a second box titled "pair programming" including the text "working in the presence of another individual on the same task (remote or in-person)". This is followed to the right by the text "impact" with an arrow to the right, final box titled "developer productivity" including text saying "satisfaction and well-being, communication and collaboration, efficiency and flow, burnout". Below this box is the text "broader impact on software development (SPACE framework)".}

  \captionsetup{justification=centering}
  \caption{Flow of ADHD characteristics and barriers mitigated by copresence practices (body doubling, pair programming) that then impact aspects of developer productivity}
  
  \label{fig:discussion-chart}
\end{figure}

We synthesize our empirical findings to answer our three primary research questions through theoretical implications in ~\S\ref{subsec:impact_copresence_theory}, implications for developer productivity in ~\S\ref{subsec:impact_space_framework}, and design implications for future copresence tools in ~\S\ref{subsec:impact_assistive_tech}. Regarding our first research question (RQ1) on how software engineers with ADHD engage in copresence, we find that while collaborative practices like pair programming broadly benefit software engineers, copresence specifically functions as assistive support to mitigate executive dysfunction for developers with ADHD. Addressing our second research question (RQ2) on challenges and strategies of copresence practices, we extend prior CSCW literature that treats copresence primarily through physical or virtual proximity~\cite{Schroeder2002, kang_social_copresence_avatar, lee_ambient_copresence, goodwin_being_there, teruyama_spatial_copresence, goebbels} through interviews of 14 developers with ADHD navigating trade-offs between cognitive support and workplace barriers. Finally, to answer our third research question (RQ3) on designing AI-assisted copresence, we provide design implications and theoretical extensions across ~\S\ref{subsec:impact_copresence_theory}, \S\ref{subsec:impact_space_framework}, and \S\ref{subsec:impact_assistive_tech}.

\subsection{Theoretical Implications Related to ADHD}
\label{subsec:impact_copresence_theory}

Our findings align with and build on aspects of copresence theory, including mutual attention, mutual behavior, and mutual emotion. Mutual attention requires two or more actors to maintain direct, reciprocal focus on one another~\cite{CamposCastillo2013}. This typically relies on continuous visual or auditory signals, such as open cameras and unmuted microphones in remote settings or physical proximity in shared workspaces~\cite{Schroeder2002}. However, our findings show that for developers with ADHD, continuous monitoring often triggers job performance-related anxiety and a state of hypervigilance. Based on our findings, we propose \textit{ambient mutual attention}, where developers can achieve a sense of mutual attention through low-fidelity, ambient signals such as soft background chatter or passive status widgets within an IDE. These ambient cues provide a sense of shared presence to help developers initiate and sustain task focus without overwhelming working memory or forcing defensive impression management across copresent peers~\cite{lampinen_copresence_social_network}.

Copresence theory describes mutual behavior as the notion of two individuals who mimic each other's behavior~\cite{CamposCastillo2013}. We find that developers with ADHD feel more motivated to initiate a task or continue making progress toward a task when their partner is also initiating or making progress on a work-related task. We also find a distinction between \textit{parallel mutual behavior} (body doubling) and \textit{joint mutual behavior} (pair programming). In parallel mutual behavior, each individual is less concerned about the nature of the other person's task, where joint mutual behavior requires cognitive alignment between the developers to be successful in making progress toward task completion. Joint mutual behavior requires transparent communication and documentation, which AI-based tools can increase the speed of. This communication shifts interactions from surface-level \textit{I-Awareness} (knowing what a partner is doing) to \textit{We-Awareness} (establishing shared reasoning and intentionality)~\cite{tenenberg2016awareness}. By automatically externalizing code rationale and summaries, AI tools help sustain this \textit{We-Awareness} without overloading the working memory of developers with ADHD.

Finally, mutual emotion refers to shared affective states, empathy, and emotional alignment between actors~\cite{CamposCastillo2013}. This aspect relies heavily on shared physiological responses and mutual emotional mirroring. Our findings demonstrate that developers with ADHD value mutual emotion primarily as a source of social-emotional support and validation during structured breaks or downtime, rather than during active code execution. Importantly, we observe a contrast in how mutual emotion operates across human-human and human-AI interactions. While human partners provide empathy and support, they also introduce perceived social judgment. Conversely, AI copresence partners completely lack emotional judgment, boredom, or social expectations. A lack of mutual emotion creates a sense of psychological safety by allowing developers to make mistakes, ask basic questions, or pause work without experiencing social shame or evaluative anxiety.

\subsection{Implications for Developer Productivity}
\label{subsec:impact_space_framework}

To examine how copresence practices influence the productivity of developers with ADHD, we analyze our findings through the \textit{SPACE framework} of developer productivity~\cite{space_dev_productivity}.
The SPACE framework measures developer productivity within five key dimensions of \textit{Satisfaction}, \textit{Performance}, \textit{Activity}, \textit{Communication}, and \textit{Efficiency}. The framework was developed across the fields of human-computer interaction, software engineering, and organizational psychology~\cite{space_dev_productivity}. Our analysis focuses specifically on \textit{Satisfaction and Well-being}, \textit{Communication and Collaboration}, and \textit{Efficiency and Flow}, which are represented the most within our empirical findings.

Within the SPACE framework, \textit{Satisfaction and Well-being} refer to the levels of happiness, fulfillment, and psychological safety developers experience while completing tasks and interacting with their environment ~\cite{space_dev_productivity}. Current professional settings often overlook emotional labor and exhaustion (i.e. invisible access labor ~\cite{Gustavsen2023Invisible, Wang2022Invisible}) experienced by developers with ADHD. We find that executive dysfunction of developers with ADHD can lead to task avoidance, internalized shame, and fear of falling behind. While copresence practices generally improve job satisfaction by providing social connection and accountability, certain scenarios where a developer has different knowledge or experiences from their partner (e.g. a junior developer paired with a senior developer) could increase performance anxiety and thus decrease overall satisfaction and well-being. The Zone of Proximal Development (ZPD) describes the gap between what a learner can accomplish independently and what they can achieve with guidance from a more capable partner~\cite{Anderson2013Learning, Chaiklin2003Zone, cole1978vygotsky}, when a senior developer navigates at their own fluency level, they push the interaction well beyond the junior developer's ZPD, lessening the space where learning can occur. Thus, strategies such as thorough on-boarding, workplace boundary setting, consistent documentation, or AI-based copresence can contribute to non-judgmental copresense sessions which allow all developers with ADHD to experience the benefits of copresence practices.

Furthermore, the SPACE framework describes \textit{Communication and Collaboration} as the quality and effectiveness of information sharing, coordination, and teamwork between developers of a team or teams through a shared mental model~\cite{space_dev_productivity}. We find that developers with ADHD experience friction in traditional synchronous collaboration and communication due to environmental barriers such as unclear requirements, unwritten social norms, or non-dynamic meeting hours. This is particularly relevant in AI-assisted development, where Storey introduces \textit{intent debt}, the loss of explicit rationale and goals, and \textit{cognitive debt}, which erodes shared mental models across developers. In human-human copresence sessions where each human is using AI, rapid code generation by one partner without real-time explanation hinders context and lessens communication and collaboration between developers, increasing the risk of intent debt~\cite{storey2026technical}. 

Finally, \textit{Efficiency and Flow} in the SPACE framework capture a developer's ability to maintain focus, minimize friction, and sustain momentum without incurring severe cognitive fatigue ~\cite{space_dev_productivity}. For developers with ADHD, the flow state (hyperfocus)~\cite{harris2017neurocognitive, gold2020review, ritonumi_flow_devs} is a strength that allows them to spend extended periods on difficult or cognitively challenging tasks. When developers enter a flow state, they reach total immersion of a particular concept and thus are able to be efficient, productive, and tend to experience a sense of enjoyment ~\cite{muller_flow, calais_flow}. However, environmental barriers such as unpredictable workplace interruptions, rigid time-tracking constraints, or latency during AI generation frequently disrupt this state, making context recovery more difficult. Developers with ADHD are able to reach flow state within body doubling and pair programming practices, but certain modalities of these practices (e.g. partner experience or personality mismatch described in \S\ref{subsubsec:interpersonal_partner_dynamics}) increase burnout and risk the accumulation of cognitive debt ~\cite{storey2026technical}. Ultimately, we find that a combination of human-human body doubling and human-AI pair programming structures optimize efficiency and flow.

\subsection{Design Implications for Copresence Technology}
\label{subsec:impact_assistive_tech}

Translating our empirical findings into actionable implications for systems, we propose three core design implications ($D_1$--$D_3$) to facilitate human-human body doubling and human-AI copresence practices for software engineers with ADHD. Each recommendation directly targets a specific phase of task execution: task initiation (\S\ref{subsubsec:task_initiation}), maintenance of flow (\S\ref{subsubsec:maintenance_flow}), and task completion (\S\ref{subsubsec:task_completion_breaks}). We map these recommendations to our empirical findings and core dimensions of copresence theory in Table \ref{tab:design_mapping}.

\begin{table*}[t]
\small
\centering
\captionsetup{justification=centering}
\caption{Mapping of design implications to empirical findings and copresence theory}
\label{tab:design_mapping}
\begin{tabular}{|>{\raggedright\arraybackslash}p{5.5cm}|>{\raggedright\arraybackslash}p{4.2cm}|>{\raggedright\arraybackslash}p{2.8cm}|}
\hline
\textbf{Design Implication} & \textbf{Empirical Finding} & \textbf{Copresence Theory} \\ \hline
$D_1$: Tools could assist with finding copresence partners and automating session initiation to alleviate activation anxiety and fear of interrupting peers & Developers desire copresence but experience initiation anxiety, internalized shame, and fear of interrupting peers (\S\ref{subsubsec:social_evaluative_anxiety}). & Mutual Attention, Mutual Behavior \\ \hline

$D_2$: Tools could provide periodic, ambient check-ins by posing contextually related follow-up questions during task execution, reinforcing accountability without inducing evaluative surveillance or micromanagement. & Developers want to participate in copresence practices with a human or AI partner who occasionally checks in on them, but without constant surveillance (\S\ref{subsubsec:motivational_mechanisms}).

& Mutual Attention, Mutual Behavior  \\ \hline

$D_3$: Tools could shift from clock-based break timers to activity-based dynamic timers through AI prediction of context switching and generation latency, where both partners could have mutual breaks during wait periods. & Rigid, clock-based timers and wait periods from AI generation latency disrupt flow and hyperfocus (\S\ref{subsubsec:distinct_structural_modalities}, \S\ref{subsubsec:ai_task_boundaries_expertise}) & Mutual Emotion, Mutual Behavior \\ \hline
\end{tabular}
\end{table*}

\subsubsection{Task Initiation ($D_1$)}
\label{subsubsec:task_initiation}
We find that developers desire copresence sessions, but experience barriers such as initiation anxiety, internalized shame, and fear of interrupting peer workflows with manual outreach. Copresence tools could support task initiation by automating session setup ($D_1$). Instead of relying on static scheduling, tools could incorporate dynamic, context-aware matching algorithms inspired by peer-study and collaborative platforms~\cite{starto_job_alg, thanh_perfect_match}. Future partner matching algorithms could use criteria including task type, estimated time of task completion, IDE activity states, and more to asynchronously pair developers with partners working towards similar goals. This matching process could provide a sense of mutual behavior and the automation of task initiation could alleviate the social anxiety our participants expressed when reaching out to peers. However, Viduchinsky described how individuals can feel compelled to perform for the algorithm, which can be different from their own identity ~\cite{Viduchinsky_algorithmic_mirror}. Future copresence tools should continue to support user agency and well-being in the matching process through matching by qualities of task rather than qualities of the individual developer.

\subsubsection{Maintenance of Flow ($D_2$)}
\label{subsubsec:maintenance_flow}

Maintaining flow for developers with ADHD is  disrupted by context switching, AI generation wait periods, and the cognitive overload of mismatching partner expertise. To lessen internal distractions without inducing performance anxiety, future copresence tools could use non-intrusive social contract reminders such as private, gentle check-ins—to reinforce accountability when developers stray off task. This could be similar to Lee et al.'s "Surrogate Avatar" which dynamically adapts its positioning in the background using a distributed optimization framework to balance real-time responsiveness with computational efficiency, ensuring presence without disrupting flow ~\cite{lee_ambient_copresence}. In pairing with mismatched experiences (e.g. junior and senior developer pairs), AI-based tools could provide background summaries and automated documentation (similar to Zoom's AI note taker ~\cite{zoom_ai_notes}) to support junior developer's learning and understanding without requiring senior developers to spend additional time on note taking and documentation.

\subsubsection{Task Completion and Breaks ($D_3$)}
\label{subsubsec:task_completion_breaks}
We find that clock-based timers with rigid schedules and wait periods from AI generation disrupt developer’s flow and states of hyperfocus.
However, identifying optimal opportunities for systems to dynamically recommend breaks was constrained by the inability to predict task boundaries or context switches. As the developer's role shifts from active coding to monitoring AI agents and managing AI-led task decomposition, predicting these state transitions to dynamically schedule contextual breaks becomes increasingly feasible. Future copresence technology could replace existing rigid timers with dynamic, ambient timers on automated, synchronized break schedules. This could be an extension of existing timer-based focus tools such as the Flow app or FocusMate ~\cite{flow_app, focus_mate, luo_time_for_break}. Kwa et al. demonstrated that while short-level inference latency of AI agents is increasingly predictable, recent benchmark evaluations demonstrate that state-of-the-art AI agents have highly variable execution time horizons ~\cite{kwa2026measuringaiabilitycomplete}. Our findings suggest that predicting AI generation wait times could mitigate workflow pauses, and future copresence tools could synchronize these wait times through dynamic breaks. Future tools could deploy break prompts, where both partners can agree to initiate a break, where the tool could then automatically unmute audio and camera streams and provide suggestions for shared discussion prompts. This structures necessary social and emotional support periods while reducing social anxiety surrounding conversation initiation. By supporting transitions into social-emotional breaks, future tools could assist with preventing developer burnout.

\subsection{Limitations}
\label{subsec:limitations}

Our findings reflect semi-structured interviews with 14 software developers with ADHD, focusing on their lived experiences with copresence practices and AI-assisted collaboration. Software engineers with ADHD represent an understudied and difficult-to-reach population in HCI research due to corporate non-disclosure agreements, privacy concerns surrounding ADHD disclosure, and high compensation levels that make research participation difficult to incentivize. Although our sample included both early-career and senior developers in different workplace settings (e.g. hybrid, remote, in-person) our findings may not fully capture the strategies of all developers with ADHD. Additionally, while our sample included diverse gender identities, larger demographic samples are needed to analyze gendered dimensions in depth. Despite these recruitment constraints, our qualitative sample achieved thematic saturation across extensive transcript data. While our objective was deep empirical understanding rather than statistical generalization, future work should evaluate copresence practices across larger sample sizes in varying settings.
\section{Conclusion}
\label{sec:06-conclusion}

In this paper, we investigate how software engineers with ADHD navigate, adapt to, and benefit from copresence practices including body doubling, pair programming, and agentic human-AI co-creation within modern hybrid workplace settings. Through semi-structured interviews with 14 software engineers with ADHD, we describe how environmental barriers and factors such as performance anxiety impact executive function, focus, and overall developer productivity. By grounding our empirical findings in copresence theory and the SPACE framework of developer productivity, we discuss how non-evaluative, AI-based tools can mitigate social anxiety, preserve shared mental models, and reduce cognitive load without disrupting developer flow. Ultimately, our findings reveal that copresence is not a one-size-fits-all construct, as developers with ADHD thrive when supported by flexible, ambient, and non-judgmental methods of collaboration. By translating these insights into design implications for future agentic copresence technology, we hope to inspire collaborative systems that foster a more inclusive and accessible software engineering workplace that expands beyond neurotypical abilities. 

\bibliographystyle{ACM-Reference-Format}
\bibliography{references.bib}

\end{document}